\documentclass[doublecol]{epl2} % for 2 columns style with line numbers
\usepackage{graphicx}
\usepackage{subcaption}

\title{Bimodality and Scaling in Recurrence Networks from ECG data}
\shorttitle{Recurrence Networks from ECG data} %Insert here a short version of the title if it exceeds 70 characters

\author{Sneha Kachhara\inst{1}\and {G. Ambika\inst{1}\footnote{g.ambika@iisertirupati.ac.in (corresponding author)}}}
\shortauthor{S. Kachhara \etal}

\institute{                    
  \inst{1} Indian Institute of Science Education and Research (IISER) Tirupati, Tirupati-517507, India \\
}

\abstract{
Human heart is a complex system that can be studied using its electrical activity recorded as Electrocardiogram (ECG). Any variations or anomalies in the ECG can indicate abnormalities in the cardiac dynamics.
In this work, we present a detailed analysis of ECG data using the framework of recurrence network (RN). We show how the measures of the recurrence networks constructed from ECG data sets, can quantify the complexity and variability underlying the data. Our study shows for the first time that the RN from ECG show the unique feature of bimodality in their degree distribution. We relate this to the complex dynamics underlying the cardiac system, with structures at two spatial scales. We also show that that there is relevant information to be extracted from the scaling of measures with recurrence threshold $\varepsilon$. Thus we observe two scaling regions in the link density for ECG data which is compared with scaling in RNs from standard chaotic and hyperchaotic systems and noise. While both bimodality and scaling are common features of RNs from all types of ECG data, we find disease specific variations in them can be quantified.}
\begin{document}

\maketitle

\section{Introduction}
The complex systems occuring in nature are mostly studied using observational or measured data over time and the methods of nonlinear time series analysis are very useful in analysing the data to understand their complexity\cite{kantz2004nonlinear}. However, for short data sets, recent technique of recurrence networks  is found very effective\cite{donner2010recurrence,zou2018complex}. In this study we use the approach of recurrence networks to study the complexity of heart dynamics  using the clinical Electrocardiogram (ECG) data of one minute duration. Such a study can bring out variations or anomalies in the ECG that carry information about the abnormalities in the underlying cardiac dynamics. 

Recently attempts have been made to understand ECG signals from a dynamical systems perspective establishing low dimensional chaos in the cardiovascular system \cite{sharma2009deterministic}. Also, significant advances have been made in the direction of cardiac modelling\cite{kotani2005model,mcsharry2003dynamical} which can generate artificial signals mimicking features of a real ECG.
Most of the studies reported in this direction are on heart rate variability (HRV) data to classify cardiac abnormalities\cite{acharya2004classification,acharya2006heart} and the analysis of the full ECG waveforms remain relatively under-explored\cite{acharya2017application}. Using multifractal measures it is reported that, the variability in the complexity of the cardiac dynamics is less in the case of healthy subjects as compared to patients\cite{shekatkar2017detecting,valenza2017complexity}. But,the conventional techniques of nonlinear time series analysis for multifractal measures require long data sets and hence not very useful for short duration clinical ECG data. 

In this study we report the interesting features that we find in the characteristics of Recurrence Networks (RNs) from short duration clinical ECG data. RN, by construction, reflects the recurrence pattern of points of the attractor or trajectory in the phase space, reconstructed from the data. Hence the measures characterising the network such as degree distribution, clustering coefficient, average path length, link density etc. can effectively quantify the complexity and variability in the ECG signals. 

Our study includes mainly ECG data from 96 cases selected from the PTB diagnostic database\cite{bousseljot2004ptb} of Physionet\cite{PhysioNet}, within the age group of 30-75 years. Among these 33 are from healthy volunteers while the rest are classified into one of the following cases:  Bundle Branch Block (BB), Cardiomyopathy (CM), Dysrhythmia (DR) and Myocardial Infarction (MI). We construct RN from each ECG time series and study how the network measures differ for typical cases. We also study how the measures vary when the threshold used in constructing the recurrence network is varied. This can provide a finer level of characterisation of the reconstructed phase space trajectory and hence the nature of the underlying dynamics. We compare our main results from ECG data sets  with similar measures derived from data of standard chaotic and hyperchaotic systems and white noise. 

Our results for the first time bring out a novel feature of the RNs from ECG data, viz. the bimodal nature of its degree distribution. We also observe a typical scaling behaviour for link density as a function of the recurrence threshold, indicating a specific structure for the phase space attractor. While  both these features are observed in general for RNs from all types of ECG data, we show their quantifiers can be useful as distinguishing measures to detect disease cases from healthy. This is because they can capture the information about any abnormalities or variations in the complexity of the underlying dynamics among the datasets.

\section{Construction of Recurrence Networks from embedded phase space attractors of ECG data}
\label{phase space}
The first step in our analysis is to reconstruct the phase space attractor from the ECG data using the method of time delay embedding\cite{bradley2015nonlinear}. As is well known, the ECG recordings are prone to noise and artefacts due to body movement, power-line interference, baseline wander and muscle contraction etc\cite{acharya2007advances}. So we do pre-processing of data to remove artefacts and trends by passing them through a band pass filter (0.5-50 Hz)\cite{berkaya2018survey}. The filtered time series is then normalized to the interval [0,1] \cite{shekatkar2017detecting}. The data thus obtained has 60000 points, which is down-sampled to 6000 points ($x_{t}$) by binning. We make sure that the whole process retains all the essential features of the original data sets.

\begin{figure}
  %\centering
  \begin{subfigure}{0.5\linewidth}
    \includegraphics[width=\linewidth]{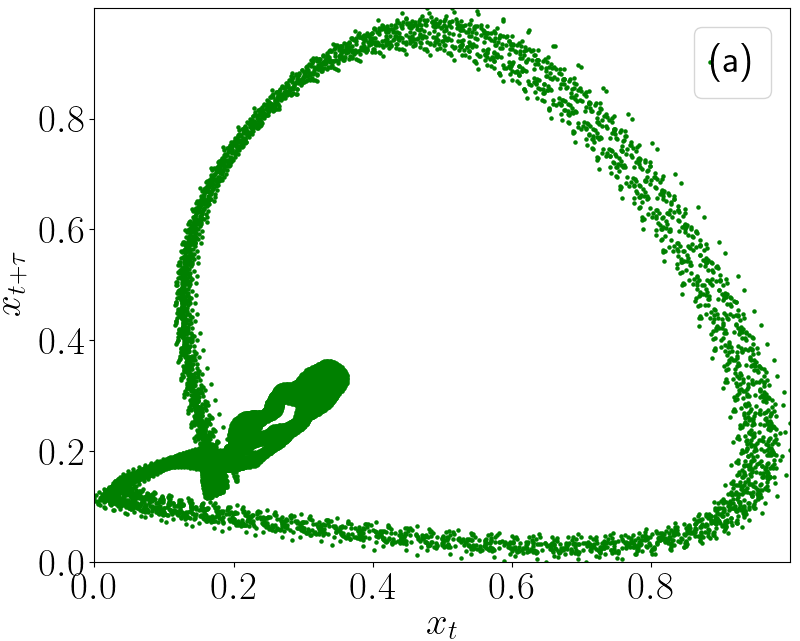}
    %\caption{Healthy}
  \end{subfigure}\hfill
  \begin{subfigure}{0.5\linewidth}
    \includegraphics[width=\linewidth]{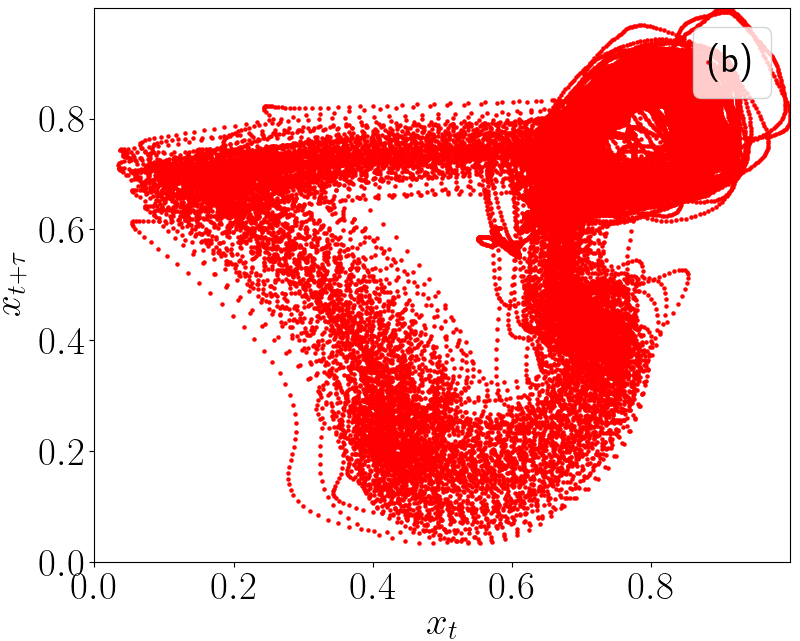}
    %\caption{BB}
  \end{subfigure}\hfill
  \begin{subfigure}{0.5\linewidth}
    \includegraphics[width=\linewidth]{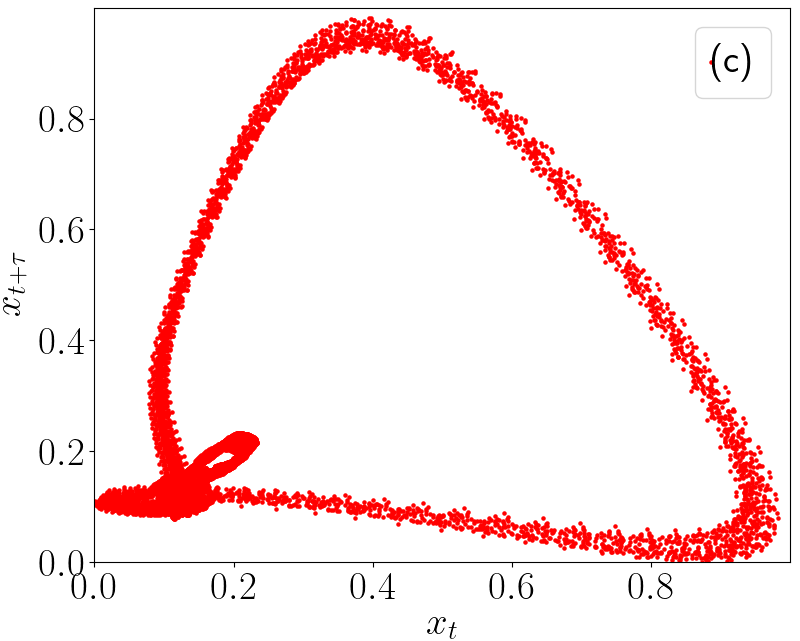}
    %\caption{CM}
  \end{subfigure}\hfill
  \begin{subfigure}{0.5\linewidth}
    \includegraphics[width=\linewidth]{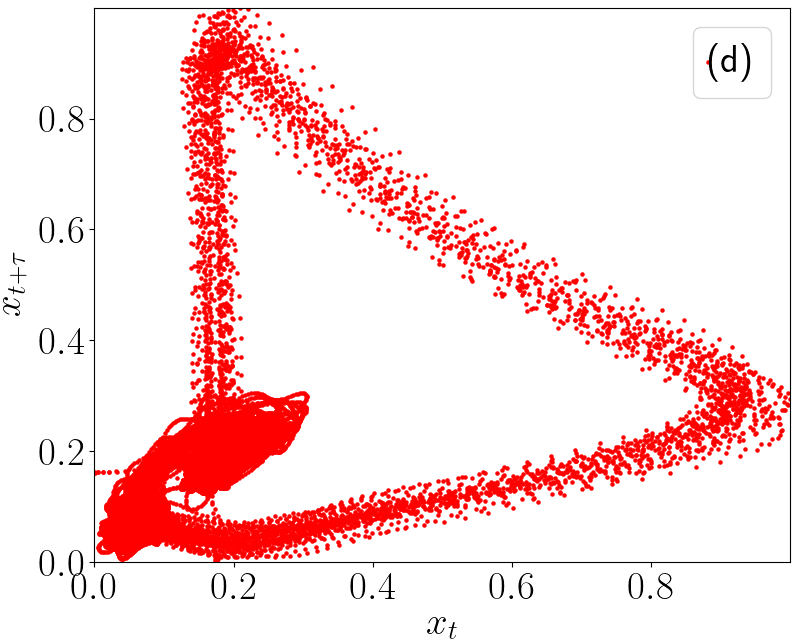}
    %\caption{DR}
  \end{subfigure}\hfill
  \begin{subfigure}{0.5\linewidth}
    \includegraphics[width=\linewidth]{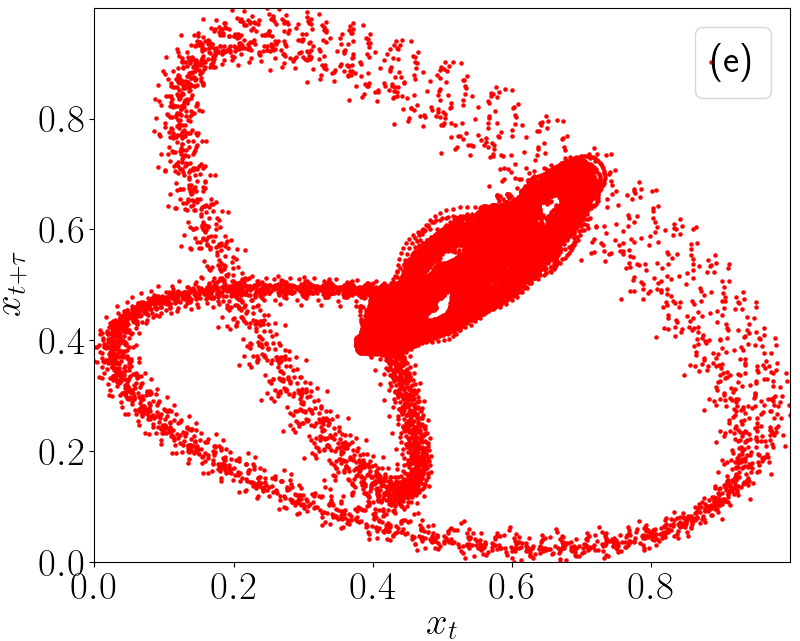}
    %\caption{MI}
  \end{subfigure}\hfill
\caption{2-dimensional projections of the reconstructed attractor in embedded space for typical ECG data sets (a) healthy subject, (b) Bundle Branch block(BB) (c) Cardiomyopathy(CM) (d) Dysrhythmia(DR) (e) Myocardial Infarction(MI).}
\label{fig:attractor}
\end{figure}

Following conventional approaches, we take the time delay for embedding as the value at which the  autocorrelation of data points, falls to $\frac{1}{e}$ of its initial value. The embedding dimension $m$ is chosen using the method of False Nearest Neighbours (FNN)\cite{hegger1999practical}. We find for most of the data sets used, the FNN gives $m = 4$ or less and hence for uniformity we embed all data in a space of dimension $m = 4$. With the time delay $\tau$ and embedding dimension $m$ thus fixed, we get the vectors in the reconstructed phase space for each data\cite{harikrishnan2006non}. A 2-dimensional projection of the reconstructed trajectory or attractor in phase space for typical cases of ECG data are shown in Fig. \ref{fig:attractor}. For healthy ECG, we observe a structure with large and small loops. In the case of BB and DR, the attractors are much wider while in the case of CM, the small scale structure is much reduced compared to healthy. The attractor has a richer structure in case of MI. We discuss how these features relate to the network measures in the following sections.

Among the various methods reported to generate complex networks from a given time series\cite{zou2018complex}, we find the  recurrence networks\cite{donner2010recurrence} is the most suitable for the short-duration 1-minute ECG data. The RN is constructed to capture the pattern of recurrences of points or dynamical states in the reconstructed phase space trajectory. This is done by defining each point as node of the network and points that recur within a chosen threshold $\varepsilon$ are connected by links. The occurrence of links is represented in the form of a binary matrix called Adjacency matrix defined as $\displaystyle A_{ij} = R_{ij}-\delta _{ij}$, where $\displaystyle R_{ij} = \Theta (\varepsilon -\left \| \vec{v_{i}}-\vec{v_{j}} \right \|)$  and $\Theta$ is the Heaviside step function\cite{donner2010recurrence}. All the required measures representing the complexity of the network are then computed from the adjacency matrix $A$.

\begin{figure}
  %\centering
  \begin{subfigure}{0.5\linewidth}
    \includegraphics[width=\linewidth]{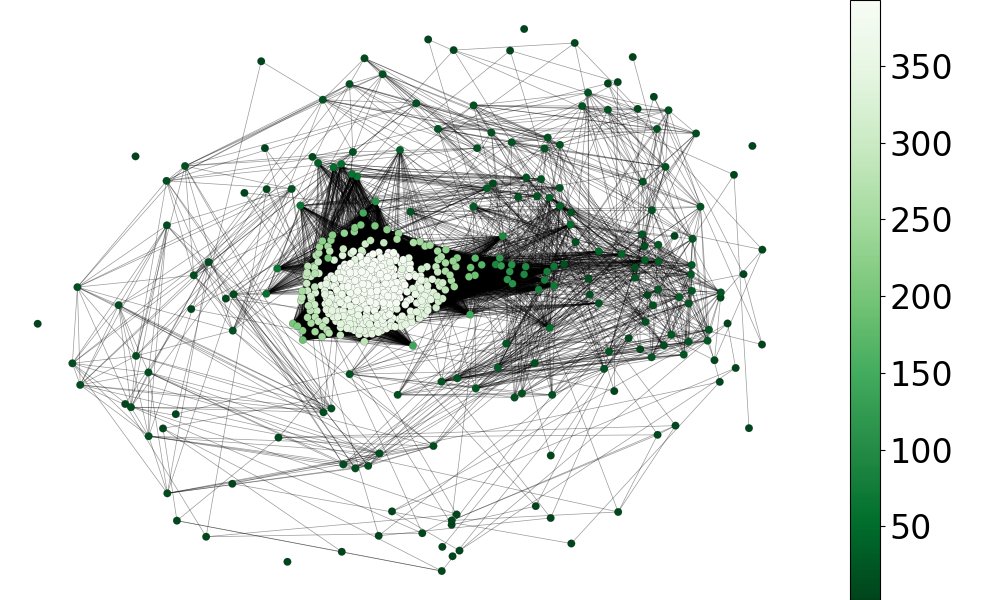}
    \caption{}
  \end{subfigure}\hfill
  \begin{subfigure}{0.5\linewidth}
    \includegraphics[width=\linewidth]{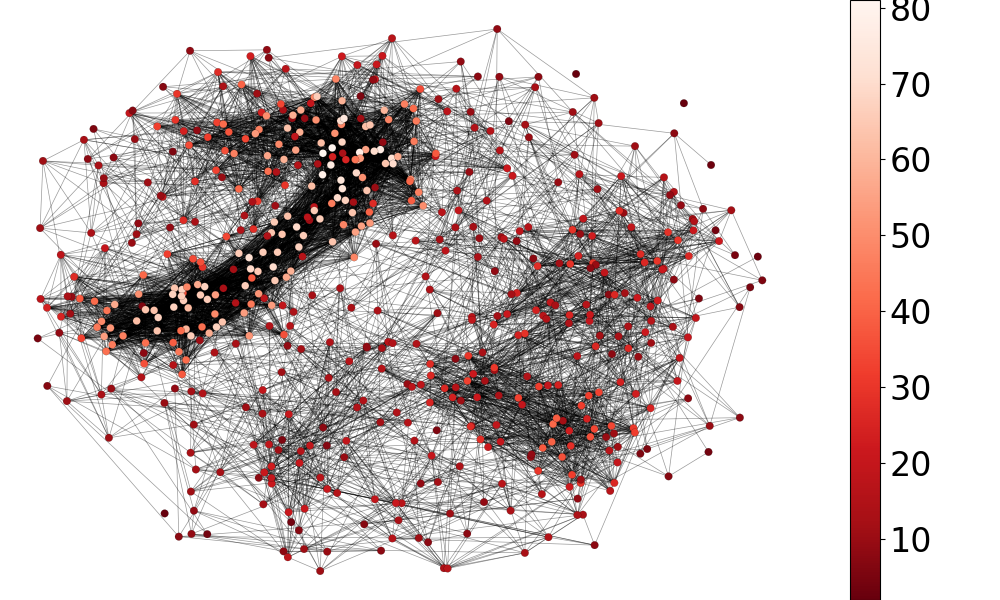}
    \caption{}
  \end{subfigure}\hfill
  \begin{subfigure}{0.5\linewidth}
    \includegraphics[width=\linewidth]{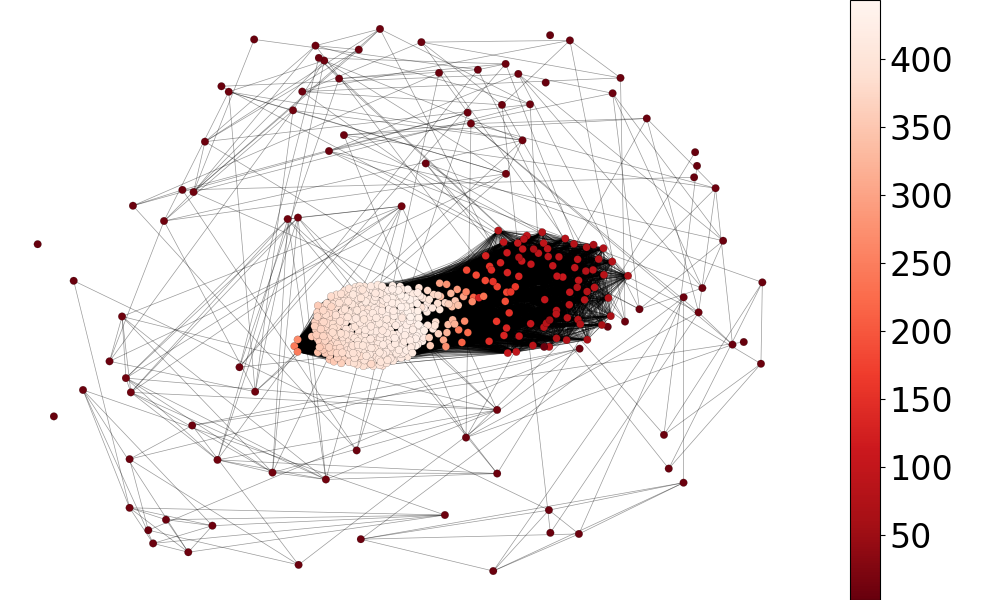}
    \caption{}
  \end{subfigure}\hfill
  \begin{subfigure}{0.5\linewidth}
    \includegraphics[width=\linewidth]{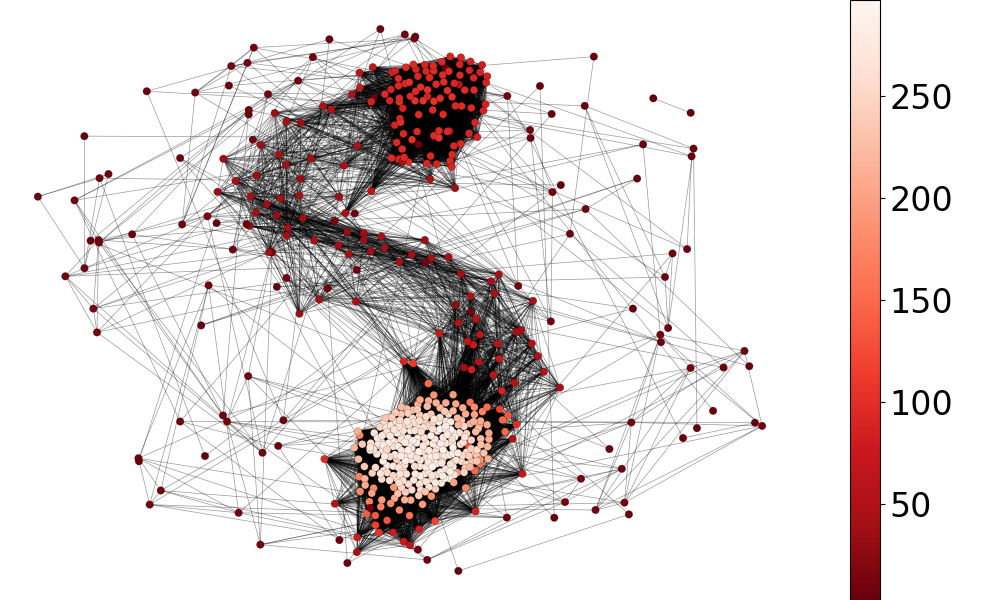}
    \caption{}
  \end{subfigure}\hfill
  \begin{subfigure}{0.5\linewidth}
    \includegraphics[width=\linewidth]{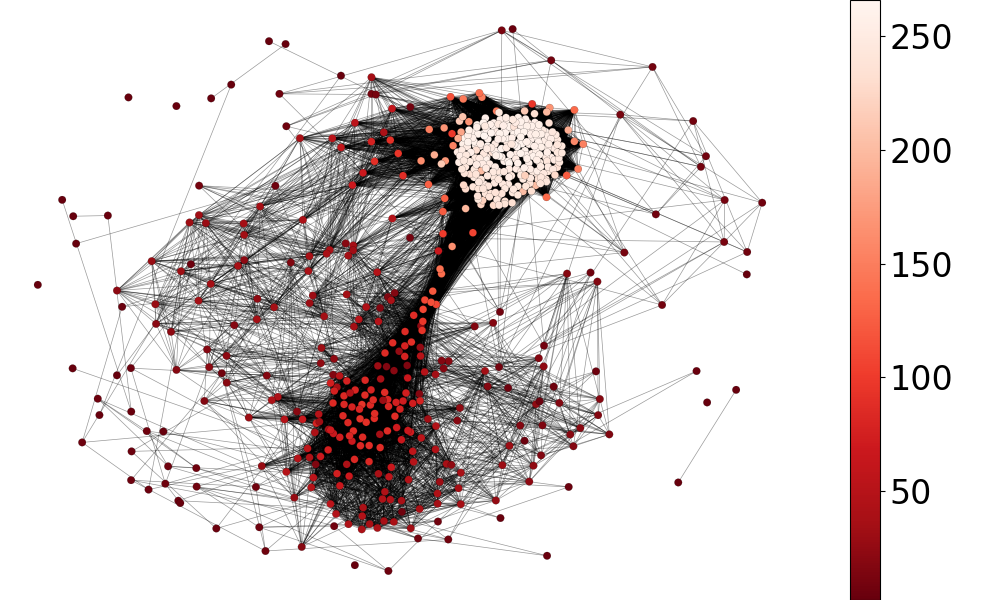}
    \caption{}
  \end{subfigure}\hfill
  \caption{Recurrence networks constructed from attractors shown in Fig. \ref{fig:attractor} for typical cases (a) healthy subject, (b) Bundle Branch block, (c) Cardiomyopathy, (d) Dysrhythmia, and (e) Myocardial Infarction. Only first 600 nodes are shown for clarity and nodes are colour coded based on their degrees.}
  \label{fig:RN_figs}
\end{figure}

The choice of $\varepsilon$ has been discussed extensively in the literature\cite{zou2018complex}. The rationale behind all of them is that the resulting network should reflect the important recurrences, but should not result in an over-connected network that will mask the relevant features. We observe that for ECG time series considered, most of the networks become just connected at $\varepsilon$ = 0.1, making it a suitable choice across all data sets. 
The RNs thus constructed for typical ECG data sets are shown in Fig. \ref{fig:RN_figs}. 

We mainly study how to characterise the RNs using  degree distribution, link density, clustering coefficient and average path length\cite{newman2010networks}. The degree of a node is defined to be the total number of its connections or links. For a given node $i$, we calculate degree $k_{i}$ as
$k_{i}=\sum_{j=1}^{N}A_{ij}$, where $N$ is the number of nodes in the network and the degree distribution $P_{k}$, is the probability of nodes with a particular degree $k$.

The link density (LD) is computed as the ratio of actual number of links in the network versus total number of possible links using,
\begin{equation}
\label{eq.1}
LD = (\frac{1}{N^{2}})\sum_{i,j=1}^{N}A_{ij}
\end{equation}

The clustering coefficient (CC) for the network is defined to be the average of the local clustering coefficients. The local clustering coefficient for a node $i$ is defined to be the ratio of the number of triads it is part of, and the number of such possible triads including the node $i$. 
\begin{equation}
\label{eq.2}
C_{i}=\frac{\sum_{j,q}A_{ij}A_{jq}A_{qi}}{k_{i}(k_{i}-1)};\ CC = \frac{\sum_{i=1}^{N} C_{i}}{N}
\end{equation}

It is clear that every pair of nodes in the network can have one or more possible paths and the shortest path between a pair of nodes (i,j) corresponds to the minimum number of links connecting them. The average path length (APL) for the network is defined as the average over the shortest distances for every pair of nodes in the networks.
\begin{equation}
\label{eq.3}
APL = \frac{1}{N(N-1)}\sum_{i\neq j}l_{ij}
\end{equation}
where $l_{ij}$ is the shortest distance (minimum number of links) between nodes $i$ and $j$.

In the following sections, we discuss how these network measures capture the underlying dynamics of the cardiac system.

\section{Measures of Recurrence networks from ECG data} \label{nw_measures}
\subsection{Degree distribution}
\begin{figure}
  %\centering
  \begin{subfigure}{\linewidth}
    \includegraphics[width=\linewidth]{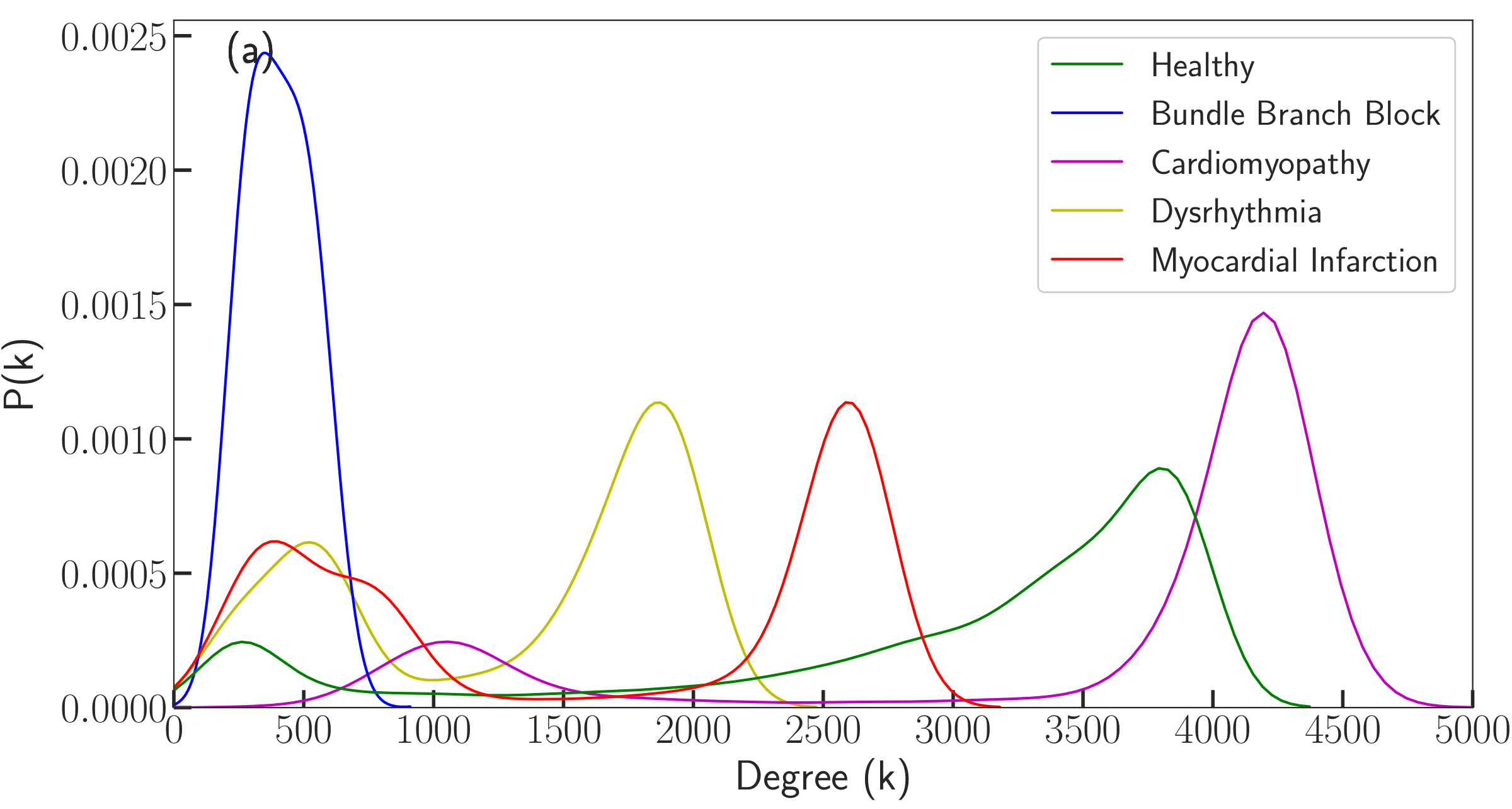}
    %\caption{All ECG}
  \end{subfigure}
  \begin{subfigure}{\linewidth}
    \includegraphics[width=\linewidth]{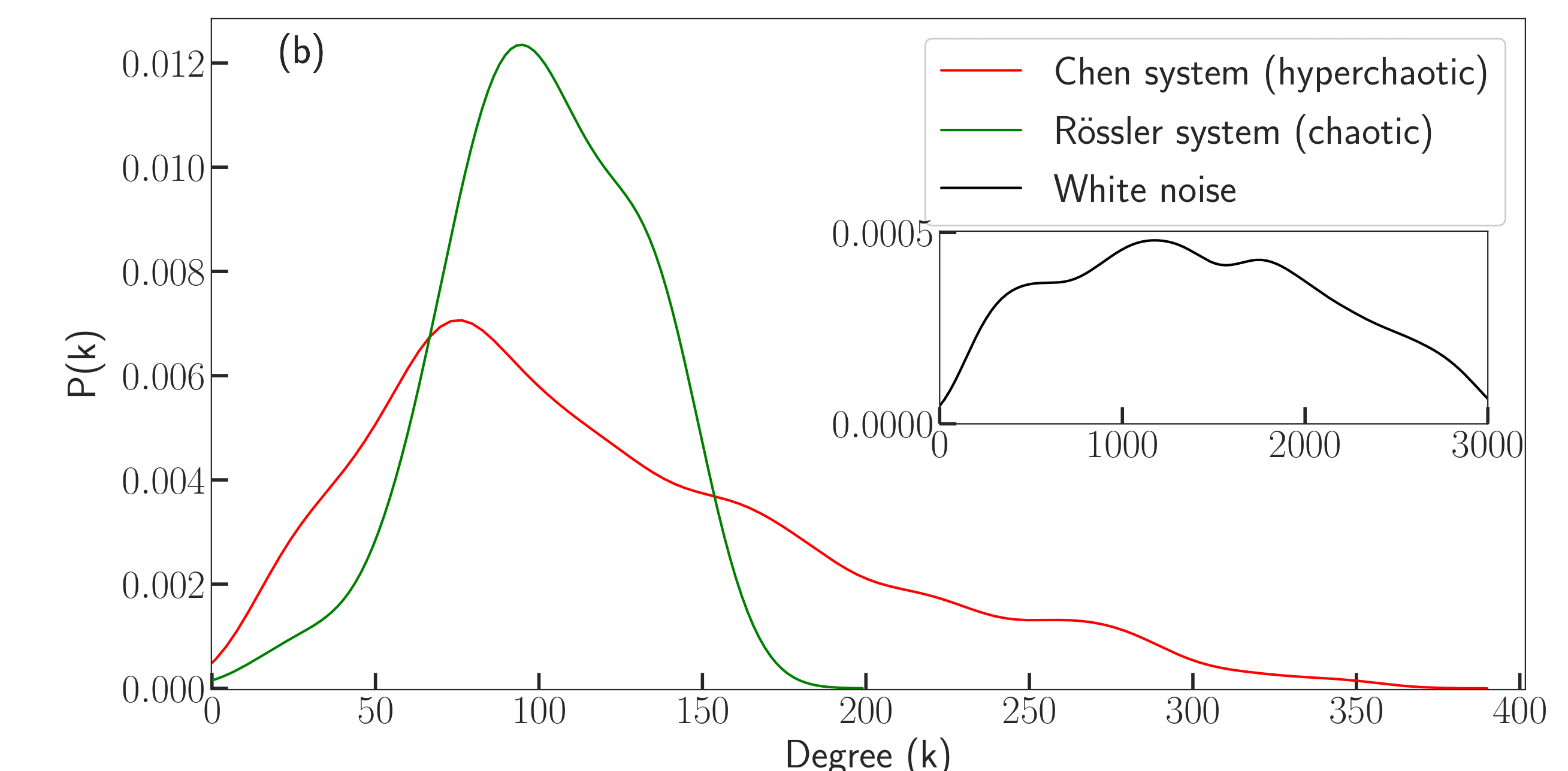}
    %\caption{Chen system, R\"{o}ssler system and White noise}
  \end{subfigure}
  \caption{Degree distributions in recurrence networks from different data sets(a)Typical cases of ECG and (b) hyperchaotic Chen system, chaotic R\"{o}ssler system and white noise. The degree distribution is bimodal in nature in all cases of ECG data sets.}
  \label{fig:deg_dist_all}
\end{figure}
The simplest measure for the structure of connections in a network is the degree distribution and complex networks are often classified using this single measure. Thus scale-free networks from real systems exhibit power law degree distribution and random networks mostly have Poisson distribution\cite{newman2010networks}. 
\begin{table*}
%\centering
\begin{center}
\begin{tabular}{ |c|c|c|c| }
\hline
Data & CC & APL & LD \\
\hline
Healthy	&	0.825	$\pm$	0.034	&	4.439	$\pm$	0.689	&	0.338	$\pm$	0.103\\
BB	&	0.741	$\pm$	0.071	&	5.364	$\pm$	2.507	&	0.199	$\pm$	0.157\\
CM	&	0.821	$\pm$	0.093	&	4.281	$\pm$	1.688	&	0.390	$\pm$	0.222\\
DR	&	0.837	$\pm$	0.044	&	4.167	$\pm$	1.134	&	0.411	$\pm$	0.119\\
MI	&	0.811	$\pm$	0.066	&	4.255	$\pm$	0.973	&	0.319	$\pm$	0.164\\
Chen	&	0.573	$\pm$	0.005	&	5.611	$\pm$	0.266	&	0.018	$\pm$	0.002\\
R$\ddot{o}$ssler	&	0.624	$\pm$	0.003	&	8.985	$\pm$	0.071	&	0.018	$\pm$	0.000\\
White Noise	&	0.477	$\pm$	0.008	&	3.213	$\pm$	0.126	&	0.034	$\pm$	0.005\\
\hline
\end{tabular}
\caption{Network measures, CC, APL and LD, for ECG data sets, standard nonlinear systems and white noise.}
\label{table: nw_measures}
\end{center}
\end{table*}
For RNs from ECG data, we observe a novel feature, bimodality in the distribution for degrees and to the best of our knowledge this has not been reported so far for recurrence networks derived from time series. The results presented are from channel-2 data from each ECG and we find that the other channels also lead to qualitatively similar results.
The presence of two peaks with almost negligible points in between seems to be a common feature of RNs from all types of ECG data.  But the distribution appears to be distorted for data from unhealthy cases compared to healthy. For comparison we also show the degree distributions from RNs of standard nonlinear dynamical systems, R\"{o}ssler system (chaotic) and the Chen system (hyperchaotic)\cite{chen2007novel}, and that from white noise, which are all unimodal in nature. 

For RNs from healthy data, the bimodal nature can be correlated to the phase space structure in Fig. \ref{fig:attractor}, where we note the presence of a small dense loop like region and a large ring. The presence of the dense region is what gives rise to the second peak at higher degrees. In the case of CM, the attractor has a highly dense small region with a large ring. For BB, the distribution is unimodal at this threshold. This must be due to the absence of small structure and a wider ring structure in phase space. The degree distribution for DR is bimodal. For MI, the first peak is broad and the distribution is bimodal.

In comparison, for white noise, the distribution is a single wide peak, indicating uniform filling of the attractor, without any small scale structures. In the case of a standard chaotic system like R\"{o}ssler system, we observe the average degree to be much lower than that of the noise and the distribution is unimodal. This behaviour can be understood in terms of the distribution of phase space points confined to a smaller region in phase space. For Chen system, in hyperchaotic region, we observe a fat tailed distribution, with a wider range of degrees but unimodal character which indicates absence of a well defined two-scaled structure in the attractor, in contrast to the ECG. Thus we see that the finer structures and complexity of the attractor is reflected in the degree distribution and in that respect, the RNs from ECG form a different class of networks.

We compute values of CC, APL and LD from RNs of all the data sets and present the results in Table \ref{table: nw_measures}. We observe that in general, the RNs from ECG have a high CC (ranging from 0.65 to 0.95) and lower APL (ranging from 3 to 8) which sets them apart from RNs of chaotic and hyperchaotic systems. We observe that the values of LD are highest for DR and lowest for BB among all ECG. Since the corresponding values of APL are lowest and highest respectively, it suggests that the RNs are more connected for DR and less so for BB. For healthy, high values of CC and APL are observed which may indicate more complexity than unhealthy cases. Moreover, the values of all measures for healthy data are marked by least variability among cases. 

\section{Scaling of measures with recurrence threshold}
As mentioned, the recurrence threshold $\varepsilon$ is crucial in the construction of RNs. In usual practice, it is chosen as the minimum value that makes all the nodes connected as a single network and subsequent analysis is performed on the RN thus constructed. It is clear that increasing $\varepsilon$, will make the network more and more connected. We conjecture that how the network measures change as $\varepsilon$ is varied, can provide a finer level of characterisation, which is not studied so far. We show we can derive relevant information about the nature of the embedded attractor and hence the data, from the nature of variation or scaling of the network measures with increasing $\varepsilon$.

\subsection{Variation of degree distribution with recurrence threshold}
\begin{figure}
  \centering
  \begin{subfigure}{\linewidth}
    \includegraphics[width=\linewidth]{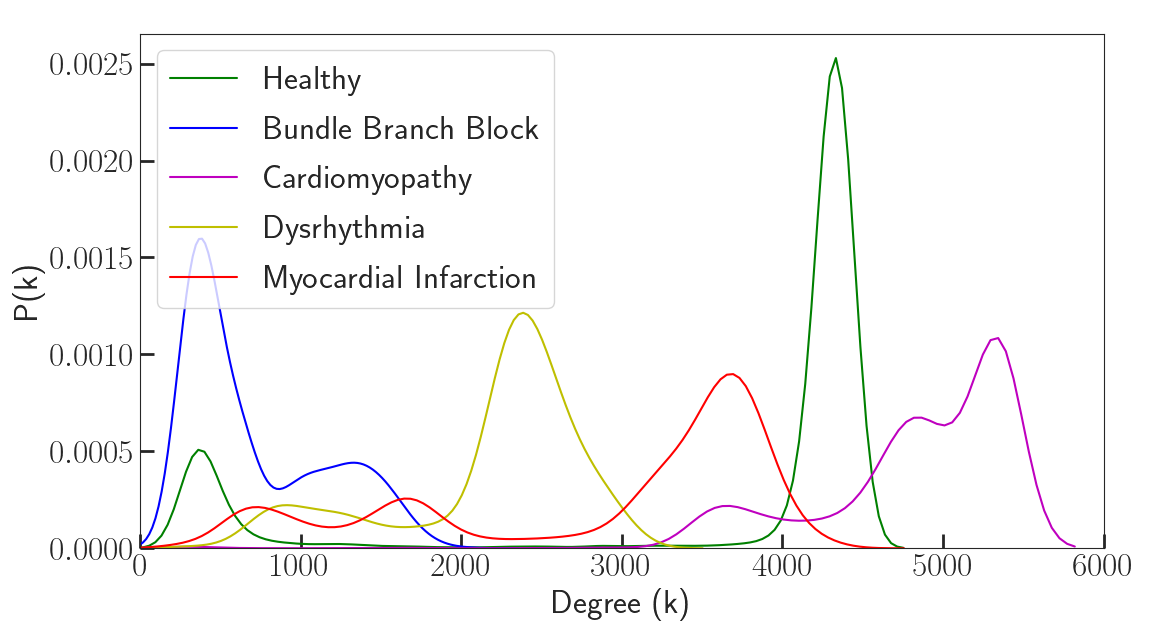}
    \caption{Degree distributions at $\varepsilon$ = 0.2}
  \end{subfigure}
  \begin{subfigure}{\linewidth}
    \includegraphics[width=\linewidth]{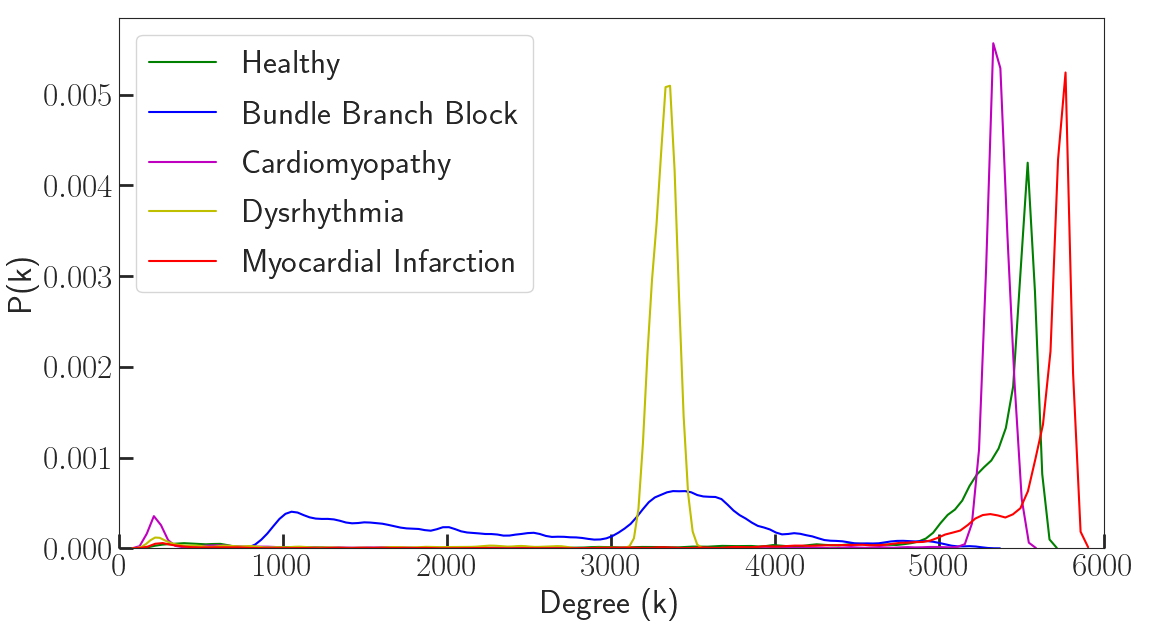}
    \caption{Degree distributions at $\varepsilon$ = 0.5}
  \end{subfigure}
  \caption{Variation in degree distribution with $\varepsilon$ for different classes of ECG data sets.}
  \label{fig:var_deg_dist}
\end{figure}
We vary recurrence threshold from 0.1 (when the networks become just connected) to 1.0, in steps of 0.01. However, we note that there is no significant change after 0.5 and hence we present the results till 0.5 ( Fig.\ref{fig:var_deg_dist}).
In the case of healthy, we observe the bimodal distribution maintains its character but first peak shifts to higher values as we increase $\varepsilon$ while the second peak becomes narrower and increases in height, in addition to shifting to higher degrees. This behaviour tells us that there are at least two spatial scales involved in the underlying dynamics.

In the case of BB, we observe a unimodal distribution at $\varepsilon$ = 0.1, with a single peak (Fig. \ref{fig:deg_dist_all}). However, as we increase the threshold to 0.2, a second peak emerges (Fig. \ref{fig:var_deg_dist}). For data from CM, we observe a behaviour similar to healthy but the distribution at higher degree saturates earlier. The RNs for data of DR in general show a wider degree distribution. As $\varepsilon$ is increased the second peak becomes narrower but maximum degree is very less as compared to other cases. For RNs from data of MI, the first peak disappears as we increase $\varepsilon$ to 0.5. 
We have presented the typical behaviour observed for each disease case as distinguished from normal cases. However, because of the presence of secondary diagnoses, some data sets show variability from this typical behaviour reported.

\subsection{Scaling of link density with recurrence threshold}\label{var_with_ld}
\begin{figure}
\begin{center}
\centering
  \includegraphics[width=\linewidth]{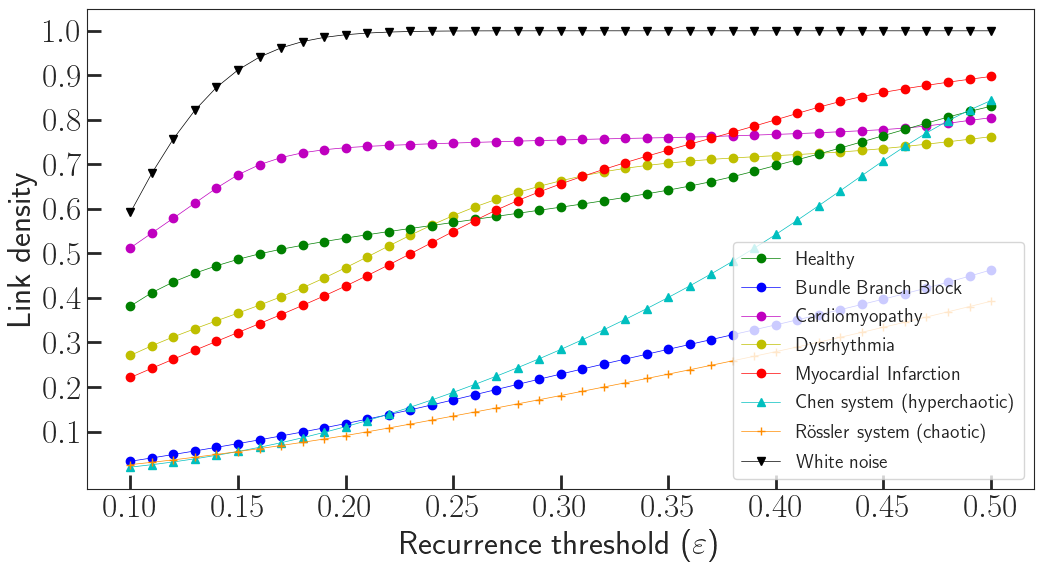}
  \caption{Variation of link density with recurrence threshold for different cases. The threshold is varied from 0.1 to 0.5.}
  \label{fig:var_with_e_ld}
 \end{center}
\end{figure}
The changes in the structure of RN as $\varepsilon$ is varied, can be studied using the link density.  This variation  as $\varepsilon$ in the range 0.1 to 0.5 is shown in Fig. \ref{fig:var_with_e_ld}. In general, there exists two scaling regions, corresponding to ranges with two slopes. But the values of slopes and the range of the two scaling regions vary among data sets. The two scaling regions are very pronounced in the case of CM while for MI, there is a gradual change of slope throughout and it is difficult to discern the distinct scaling regions. For BB, on the other hand, the link density is much smaller and there is almost no change of slope. This corresponds to a simple attractor which is very similar to the chaotic R\"{o}ssler system. In the case of DR the first scaling region is apparent from 0.1 to 0.3, but the second one appears in the range 0.35 to 0.5, corresponding to a sparser attractor with no intermediate structure.

For comparison, we also present results from white noise, R\"{o}ssler system and Chen system. For R\"{o}ssler system, we observe an almost linear variation, while for Chen system we observe two scaling regions. However, the change in slope is gradual and the value of link density for smaller thresholds is low, indicating that the regions in the attractor on a small scale are less dense compared to ECG data. The link density for white noise, on the other hand, rises very sharply with $\varepsilon$ and saturates at a very small value of around 0.2, as expected. 

The presence of two scaling regions indicates typical structure for underlying attractor of ECG data and the differences in values of slopes can be an indicator to distinguish different cases of disease from healthy ECG. For this, we take the scaling of LD with $\varepsilon$  as LD $\sim$ $\varepsilon^{\gamma}$ and compute values of $\gamma$ from their log-log plots. The results for the indices $\gamma_{1}$ and $\gamma_{2}$ corresponding to the scaling regions are given in Table \ref{table:slopes}.
We would like to emphasise that although the behaviour of ECG data shown in Fig. \ref{fig:var_deg_dist} and \ref{fig:var_with_e_ld} are shown for typical cases from each type, the results in tables \ref{table: nw_measures} and \ref{table:slopes} are average values computed over all data sets. For standard systems, the values are averaged over data generated from different initial conditions. The presence of secondary diagnosis obscures some results and makes it difficult to separate disease specific features. However, the trends on an average can still be picked up and can guide further, detailed analysis of the time series. 

\begin{table}
\centering
\begin{tabular}{|c|c|c|}
\hline
Data & Scaling index $\gamma_{1}$ & Scaling index $\gamma_{2}$ \\
\hline
Healthy	&	0.782	$\pm$	0.288	&	0.387	$\pm$	0.296\\
BB	&	1.248	$\pm$	0.509	&	0.830	$\pm$	0.364	\\
CM	&	0.891	$\pm$	0.621	&	0.407	$\pm$	0.401	\\
DR	&	0.609	$\pm$	0.271	&	0.426	$\pm$	0.368	\\
MI	&	0.941	$\pm$	0.449	&	0.372	$\pm$	0.276	\\
Chen	&	2.462	$\pm$	0.059	&	2.136	$\pm$	0.199\\
R$\ddot{o}$ssler	&	1.859	$\pm$	0.010	&	1.263	$\pm$	0.017\\
White Noise	&	1.451	$\pm$	0.157	&	0.000	$\pm$	0.000\\
\hline
\end{tabular}
\caption {Scaling indices for link density calculated from log -log plots for ECG data of all types, standard nonlinear systems and white noise. \label{table:slopes}}
\end{table}

\section{Conclusion}
In this study, we construct recurrence networks from clinical ECG data of 1-minute duration and analyse their measures for characterisation. We also compare them with those from the standard nonlinear dynamical systems of R\"{o}ssler (chaotic) and Chen (hyperchaotic), and white noise. The RNs from ECG are found to be different from standard nonlinear dynamical systems and noise in significant ways. The degree distribution of RNs from ECG exhibits a bimodal character and thus form a separate class. We note that this is the first time this novel feature is reported for RNs from time series. We reason that the complex dynamics underlying cardiac system, with structures at two spatial scales for the phase space attractor, is revealed through this unique feature.

RNs from ECG in general have significantly higher value for the clustering coefficient and lower value for the average path length. Among different cases of ECG, Bundle Branch block has values for CC and APL that can be identified separately from all other classes. The healthy show least variability in both measures, and on average have a higher value of both CC and APL.

Our study also indicates that there is relevant information to be extracted from the behaviour of measures as recurrence threshold is varied. In particular, we observe two scaling regions in the link density vs $\varepsilon$ curve for ECG data, that is very different from data of standard chaotic systems and white noise. The scaling indices are calculated for all data sets to quantify these characteristics.
The study on how the measures vary as we change the threshold, characterise the dynamics at a finer level. This is the first time such variations are analysed to advantage in the classification of data sets. The analysis presented is to be applied to larger number of data sets so that disease specific ranges of measures and scaling indices can be derived. This work is already in progress and the results will be published elsewhere soon.

The present study closely resembles the tools of multifractal analysis from data in many respects. But in addition to being easier to implement, it has the specific advantage that it requires only short data sets for its implementation. This makes the method faster for practical purposes and more effective for short and nonstationary data and hence can be used for quantifying changes with time for data from the same source.

\section{Acknowledgements}
One of the authors, Sneha Kachhara acknowledges financial support from Council of Scientific and Industrial Research (CSIR), India. We thank Prof. K.P. Harikrishan for useful discussions.

\bibliographystyle{eplbib}

\begin{thebibliography}{10}
\expandafter\ifx\csname url\endcsname\relax\def\url#1{\texttt{#1}}\fi

\bibitem{kantz2004nonlinear}
\Name{Kantz H. \and Schreiber T.} \Book{Nonlinear time series analysis} Vol.~7
  (Cambridge university press) 2004.

\bibitem{donner2010recurrence}
\Name{Donner R.~V., Zou Y., Donges J.~F., Marwan N. \and Kurths J.} \REVIEW{New
  Journal of Physics}{12}{2010}{033025}.

\bibitem{zou2018complex}
\Name{Zou Y., Donner R.~V., Marwan N., Donges J.~F. \and Kurths J.}
  \REVIEW{Physics Reports}{}{2018}{}.

\bibitem{sharma2009deterministic}
\Name{Sharma V.} \REVIEW{The open cardiovascular medicine
  journal}{3}{2009}{110}.

\bibitem{kotani2005model}
\Name{Kotani K., Struzik Z.~R., Takamasu K., Stanley H.~E. \and Yamamoto Y.}
  \REVIEW{Physical Review E}{72}{2005}{041904}.

\bibitem{mcsharry2003dynamical}
\Name{McSharry P.~E., Clifford G.~D., Tarassenko L. \and Smith L.~A.}
  \REVIEW{IEEE transactions on biomedical engineering}{50}{2003}{289}.

\bibitem{acharya2004classification}
\Name{Acharya R., Kumar A., Bhat P., Lim C., Kannathal N., Krishnan S. \etal}
  \REVIEW{Medical and Biological Engineering and Computing}{42}{2004}{288}.

\bibitem{acharya2006heart}
\Name{Acharya U.~R., Joseph K.~P., Kannathal N., Lim C.~M. \and Suri J.~S.}
  \REVIEW{Medical and biological engineering and computing}{44}{2006}{1031}.

\bibitem{acharya2017application}
\Name{Acharya U.~R., Fujita H., Oh S.~L., Hagiwara Y., Tan J.~H. \and Adam M.}
  \REVIEW{Information Sciences}{415}{2017}{190}.

\bibitem{shekatkar2017detecting}
\Name{Shekatkar S.~M., Kotriwar Y., Harikrishnan K. \and Ambika G.}
  \REVIEW{Scientific reports}{7}{2017}{15127}.

\bibitem{valenza2017complexity}
\Name{Valenza G., Citi L., Garcia R.~G., Taylor J.~N., Toschi N. \and Barbieri
  R.} \REVIEW{Scientific reports}{7}{2017}{42779}.

\bibitem{bousseljot2004ptb}
\Name{Bousseljot R., Kreiseler D. \and Schnabel A.} \REVIEW{physionet.
  org}{}{2004}{}.

\bibitem{PhysioNet}
\Name{Goldberger A.~L., Amaral L. A.~N., Glass L., Hausdorff J.~M., Ivanov
  P.~C., Mark R.~G., Mietus J.~E., Moody G.~B., Peng C.-K. \and Stanley H.~E.}
  \REVIEW{Circulation}{101}{2000 (June 13)}{e215} circulation Electronic Pages:
  http://circ.ahajournals.org/content/101/23/e215.full PMID:1085218; doi:
  10.1161/01.CIR.101.23.e215.

\bibitem{bradley2015nonlinear}
\Name{Bradley E. \and Kantz H.} \REVIEW{Chaos: An Interdisciplinary Journal of
  Nonlinear Science}{25}{2015}{097610}.

\bibitem{acharya2007advances}
\Name{Acharya R., Krishnan S.~M., Spaan J.~A. \and Suri J.~S.} \Book{Advances
  in cardiac signal processing} (Springer) 2007.

\bibitem{berkaya2018survey}
\Name{Berkaya S.~K., Uysal A.~K., Gunal E.~S., Ergin S., Gunal S. \and
  Gulmezoglu M.~B.} \REVIEW{Biomedical Signal Processing and
  Control}{43}{2018}{216}.

\bibitem{hegger1999practical}
\Name{Hegger R., Kantz H. \and Schreiber T.} \REVIEW{Chaos: An
  Interdisciplinary Journal of Nonlinear Science}{9}{1999}{413}.

\bibitem{harikrishnan2006non}
\Name{Harikrishnan K., Misra R., Ambika G. \and Kembhavi A.} \REVIEW{Physica D:
  Nonlinear Phenomena}{215}{2006}{137}.

\bibitem{newman2010networks}
\Name{Newman M.} \Book{Networks: an introduction} (Oxford university press)
  2010.

\bibitem{chen2007novel}
\Name{Chen Z., Yang Y., Qi G. \and Yuan Z.} \REVIEW{Physics Letters
  A}{360}{2007}{696}.

\end{thebibliography}

\end{document}